\documentclass{article}
\usepackage[dvips,unicode]{hyperref} 
\usepackage[dvips]{graphicx} 
\usepackage{amssymb} 

\newcommand{\kT}{\kappa \! T}

\begin{document}

\begin{center}
{\Large \textbf{Stochastic Dynamics of Magnetosomes in Cytoskeleton}}
\\ ~\\
V.N. Binhi$^1$ and D.S. Chernavskii$^2$ 
\\ ~\\
$^1$A.M. Prokhorov General Physics Institute RAS, Moscow, Russia \\
$^2$P.N. Lebedev Physics Institute RAS, Moscow, Russia
\end{center}
\vspace{1mm}

\begin{abstract}
Rotations of microscopic magnetic particles, magnetosomes, embedded into the cytoskeleton and subjected to the influence of an ac magnetic field and thermal noise are considered. Magnetosome dynamics is shown to comply with the conditions of the stochastic resonance under not-too-tight constraints on the character of the particle's fastening. The excursion of regular rotations attains the value of order of radian that facilitates explaining the biological effects of low-frequency weak magnetic fields and geomagnetic fluctuations. Such 1-rad rotations are effectively controlled by slow magnetic field variations of the order of $200\mbox{\,nT}$. 
\end{abstract}
\vspace{2mm}
\vspace{4mm}

There are many hypothetical mechanisms suggested to explain the biological effects of weak low-frequency magnetic fields. A brief review of the mechanisms may be found in \cite{binhi03ae} and the detailed discussion in \cite{binhi02}. 
At the same time, the physical nature of these effects remains unclear. The basic problem is that the interaction energy of biologically active molecules and the MF at the geomagnetic level is very small \cite{binhi02b}. It is much smaller than the energy of thermal fluctuations $\kT \approx 4\cdot 10^{-14}\mbox{\,erg}$ at physiological temperatures. 

However, many organisms are well known to contain submicron magnetic particles. The energy of their turn in a weak magnetic field $H$ is substantially greater than $\kT $. For single-domain magnetite particles of radius $r = 10^{-5}\mbox{\,cm}$ or $100\mbox{\,nm}$ in the geomagnetic field the energy $\mu H \approx vJ H$ equals approximately $24 \kT $, where $\mu$ is the magnetic moment of the particle, $v$ and $J$ are the volume and the saturation magnetization. 

The cytoplasm near cell membranes features such visco-elastic properties that the turning of a microparticle may serve as a stimulus to cell division or ignite a nerve impulse. Magnetite particles found in the brain tissues of animals and humans are of particular interest: this constitutes one of the possible mechanisms of the weak MF effect on the human organism \cite{kirschvink85-e}. 
The nerve tissue of the brain is separated from the circulatory system by the blood-brain barrier which is impermeable for most chemicals. In turn, the circulatory system is separated from the digestive system. Therefore, relatively large ferro- or ferrimagnetic particles cannot penetrate into brain tissue as a pollutant. They are found to have a biogenic origin, i.e. they appear over time as a direct result of the crystallization in brain matter. Biogenic magnetite particles are often called `magnetosomes'; they were first discovered in bacteria that displayed magnetotaxis \cite{blakemore75}.

The density of magnetosomes in the human brain is more than $5\cdot 10^6$, and in meninges more than $10^8$ crystals per gram \cite{kirschvink92b}. In fact, about 90\% of the particles measured in this study were 10--$70\mbox{\,nm}$ in size, and 10\% were 90--$200\mbox{\,nm}$. The particles were grouped into ensembles of 50--100 crystals.

Given the fact that magnetic moment is in direct proportion to the particle's volume, it is easily seen that the inequality $\mu H < \kT $ is true for particles less than $30\mbox{\,nm}$ in size. Due to thermal disturbances, such particles can spontaneously switch their magnetic flux without turns, i.e. they are in a superparamagnetic state. The particles that are several hundred nm and more in size go to multiple-domain states (the energy of domain walls is less than that of the MF produced by a single-domain state): their remanent magnetization may be ignored. These particles experience almost no torque in the magnetic fields under consideration. In this article we consider the dynamics of an idealized 'mean' particle, the magnetosome with the radius $r \sim 100\mbox{\,nm}$ in a single-domain state.

The energy of the magnetosome in the geomagnetic field is $\approx 24 \kT $; when exposed to an additional variable magnetic field $h$, its regular changes are about $(h/H_{\rm geo})24kT$. If these changes exceed thermal fluctuations $\sim \kT /2$, they can cause a biological response. This sets a natural constraint on the MF magnitude capable of affecting a biophysical or biochemical system appreciably: $h \gtrsim 1$--$2\mbox{\,$\mu $T}$. However, for magnetosomes bound to an oscillator system, eigenfrequency of which is close to the frequency of the external field, the biologically detectable level of the MF might be less. This may also take place in the special case of magnetosomes bound to a visco-elastic medium: then thermal fluctuations work to facilitate rather than impede the capability of a weak magnetic stimulus to cause a response. 

Oscillations of a protein macromolecule (dipole resonator) in a microwave EMF have been studied in \cite{chernavsky89-e}. The dynamics of oscillating magnetic particles in the ELF MF has not been studied in detail yet. Theoretical evaluations of the magnetoreception mechanism based on magnetosome rotations in MF have been working out by many authors since 1970s \cite{yorke79}.

In known works, the dynamics of magnetosomes was modelled by using the equation of {\em free rotations} in a viscous liquid, since the elastic properties of structures to which magnetosomes may be attached were not assessed. Quasi-elastic torque had been considered only in relation to the magnetic moment energy in the constant geomagnetic field. It turns out that explicitly taking into account the elasticity of the medium enables one to describe a stochastic rotational dynamics of magnetosomes that may be used to explain the particularities of magnetoreception of weak and hyperweak MFs. 

This article considers the dynamics of a magnetite particle embedded in the cytoskeleton. The latter consists of a 3D net of protein fibers of 6 to $25\mbox{\,nm}$ in diameter that include actin filaments, intermediate filaments, and microtubules. The ends of these fibers may be fastened to the membrane surface and to various cell organelles. We assume the fibers may also be fastened to a magnetosome surface normally covered with a bilayer lipid membrane \cite{gorby88}. This fixes the position of the magnetosome and constrains its rotation to some extent. The stationary orientation of the magnetosome generally does not follow the constant MF direction. The balance of the elastic and `magnetic' torques determines the orientation now. The torque $\bf m$ affecting a particle of the magnetic moment $\boldmath \mu$ in an MF $\bf H$ equals ${\bf m} = \boldmath \mu \times {\bf H}$. Here, putting aside the 3D character of the magnetosome rotations, we consider the magnetosome's motion in the plane of two vectors: the unit vector $\bf n$ of the $x$-axis, with which the vector of magnetosome's magnetic moment coincides in the absence of the MF (equilibrium position, $\varphi =0$), and the MF vector $\bf H$. 
The Langevin equation for rotational oscillations of the particle is as follows:
\begin{equation} \label{01} I \ddot{\varphi } + \gamma \dot{\varphi } + k \varphi = -\mu H(t) \sin(\varphi - \varphi_0) + \xi'(t) ~,~~~ \omega_0=\sqrt{k/I}~, \end{equation} where $\varphi$ is the angular displacement, $I$ is the moment of the particle's inertia, $\gamma$ is the dissipation coefficient, $k$ is the factor of mechanical elasticity resulting from the cytoskeleton fibers' bending, $\xi'(t)$ is a stochastic torque with the correlation function $\langle \xi'(t) \xi'(t+\Delta t) \rangle = 2\gamma \kT  \delta(\Delta t)$, while $\omega_0$ is the eigenfrequency, and $\varphi_0$ is the MF direction. 
Then, we assume the quantity of fibers fastening the magnetosome to the cytoskeleton may vary from particle to particle and a significant number of magnetosomes are mobile enough to markedly change their orientation in the geomagnetic field. This means the mechanical elasticity due to the fibers' bending is of the same order as or less than the magnetic elasticity $k \lesssim \mu H \approx 24 \kT  $. For magnetite Fe$_3$O$_4$ particles with the substance density $\rho \approx 5.2\mbox{\,g/cm$^3$}$ and radius $r \sim 10^{-5}\mbox{\,cm}$, we derive a value $\omega_0$ in the order of $10^{6}\mbox{\,rad/s}$. A resonance, however, is not possible since the inertia forces are much less than viscous forces: $ I\omega_0 \ll \gamma $. Hereafter, the inertia term in the equation of motion may be ignored.

\begin{figure}
\[ \includegraphics[scale=0.5]{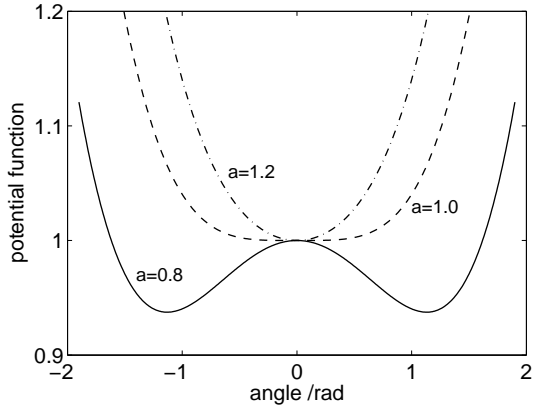} \]
\caption{Potential function of a magnetosome at different values of the elastic parameter $a$.}
\label{f.1}
\end{figure}

The idea of this work is to study the dynamics of a magnetosome fixed into a {\em visco-elastic} cytoskeleton and predominantly oriented in a direction opposite to that of a constant MF; here the case $\varphi_0 = \pi $ is considered. The potential energy of a magnetosome in terms of $\mu H$ in the absence of ac MF 
\[ U = \cos(\varphi) + \frac a2 \varphi^2 ~,~~~ a= \frac{k} {\mu H } \] is shown in Fig.\,\ref{f.1}. As is seen, for not too large angles at $a<1$ there are two stable equilibrium positions $\varphi_{\pm}$ and the unstable one $\varphi_0 =0$. 
Within each of the wells of this double-well potential, the motion of the magnetosome demonstrates no peculiarities. This sort of motion has been repeatedly considered in literature. At the same time, due to thermal disturbances, the transitions appear from well to well even with no ac MF signal. Given that, the stochastic turns of the particle take place with considerable angular displacements. A deterministic external force, the ac MF in our case, causes such transitions to be somewhat ordered, the maximum order attained just at the optimal level of the noise. It is essentially the phenomenon of the so-called stochastic resonance (SR) first introduced in \cite{benzi81} to explain some geophysical processes. 

So far, the probable manifestation of the SR in the dynamics of magnetosomes has not been investigated. Consider the joint influence of a magnetic signal and a random torque $\xi'(t)$ on a magnetosome. The equation of motion takes the form:
$ \gamma \dot{\varphi} -\mu H \sin(\varphi) +k\varphi - \mu h \sin (\Omega t) = \xi'(t)~. $ 
With the designations 
\begin{equation} \label{99} h' \equiv \frac {h} {H} ~,~~ \beta \equiv \frac {\gamma \Omega} {\mu H}~,~~ \tau \equiv \frac {\mu H} {\gamma} t~,~~ D\equiv \frac {2\kT } {\mu H} \end{equation} the equation is reduced to 
\begin{equation} \label{100} \dot{\varphi} + \partial_{\varphi} U(\varphi, \tau) = \sqrt{D} \xi(\tau) \end{equation} with the potential
\begin{equation} \label{1001}
U(\varphi, \tau) = \cos (\varphi) + \frac a2 \varphi^2 - \varphi h' \sin(\beta \tau)~ . \end{equation} Here $\xi(\tau)$ is the centered Gaussian process of unit variance (the identity $\delta(\alpha t)= \delta(t)/|\alpha | $ is used). 

Several SR theories are known; we use the results of \cite{mcnamara89}, where the general expression has been derived for the power spectrum of oscillations of a bistable system agitated by regular and random signals. The signal-to-noise ratio is determined as the ratio of the spectrum amplitude at the frequency of the regular signal, to the level of noise at the same frequency. For the system (\ref{100}) with a general double-well potential the signal-to-noise ratio equals
\begin{equation} \label{101} R_{\rm sn} = \sqrt{|U''(\varphi_0)| U''(\varphi_{\pm})} \, \frac{U_1^2} {D^2} \exp \left( -2 U_0/D \right) ~,\end{equation} where $U_0$ and $U_1$ are the height and modulation amplitude of the potential barrier, and $U''$ is the curvature of the potential at the respective equilibrium points. The function (\ref{101}) attains its maximum at the optimal level of noise $D=U_0$. This means there is an interval in the value of $D$, where the signal-to-noise ratio unexpectedly increases along with increasing noise power --- it is this that is the signature of SR. 

\begin{figure}
\[ \includegraphics[scale=0.5]{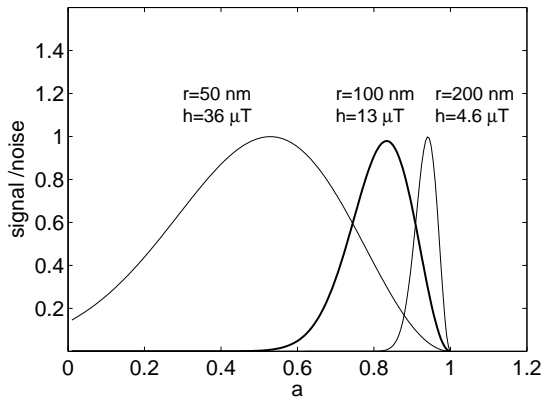} \]
\caption{Signal-to-noise ratio in rotations of the magnetosome of different radii and at different amplitudes of the ac MF.}
\label{f.2}
\end{figure}

Quantities $U_0$ and others of the potential (\ref{1001}) have no exact analytical presentations. Here we derive them as expansions over the parameter $1-a$ that is assumed to be a small parameter:
\[ \varphi_+^2 = 6 (1-a),~~ U_0 = \frac 32 (1-a)^2 ,~~ U_1^2 = 6 h'^2 (1-a) ,~~ U''(0)= a-1 ,~~ U''(\varphi_+) = 2(1-a) ~. \] Substitution into (\ref{101}) leads to the following expression for the signal-to-noise ratio:
\[ R_{\rm sn} \approx \frac {6\sqrt2 \,h'^2 (1-a)^2} {D^2} \exp\left\{-3 (1-a)^2 /D \right\} ~. \] This function is plotted in Fig.\,\ref{f.2} at various ac MF amplitudes and values of the noise parameter $D$, which depends on the size of the magnetosome.

As is seen, there is a marked interval of the elasticity parameter $a=k/\mu H$, wherein the signal-to-noise ratio is close to unity. The 100-nm magnetosome fixed in the cytoskeleton with elasticity $a= 0.7$--0.9\,$\mu H$ in the 13-$\mu$T ac MF and 46-$\mu$T geomagnetic field regularly turns at angles of the same order as the chaotic rotations. It is particularly evident for 50-nm particles, that almost all of them are in the SR conditions. 200-nm particles make regular turns at relatively small MFs $\approx 4.6\mbox{\,$\mu $T}$. Although in each of these cases there is no gain in the magnitude of the effective ac MF as compared to the case of a single-well motion, it is important that the rotation excursion is an order higher, about 1 rad. With such excursions it is easier to account for the influence of the magnetosome's rotations on biochemical processes. 

Note that in an SR, the signal-to-noise ratio is enhanced because of the reduced coherence of the signal present in the spectrum of magnetosome oscillations as compared to the coherence of the ac MF signal. Therefore MF signal detection requires a discrimination system --- probably, a nonlinear system of biochemical reactions with the characteristic time $\sim 1/\Omega$ --- which can `make a decision' as to whether a signal is present in noise. 

It is of the essence, that this primary mechanism of magnetobiological effects, displaying no frequency selectivity in the ELF range, nevertheless allows one to verify it experimentally. Since the parameter $a=k/\mu H$ depends on the constant MF, the `resonance' on Fig.\,\ref{f.2} will show itself also as a `window' in constant MF values when the effect is possible. Therefore, provided the MF signal transduction to the biochemical level is governed by an SR with magnetosomes of a certain size within limits of about 10--20{\%}, it follows that the biological effect in the ac MF will take place only in a constant MF near the level $H\sim k/a\mu$. Indeed, when the MF decreases, the potential function transforms into a single-well one and large rotational excursions are no longer possible. When the MF increases, the potential barrier grows and the magnetosome finally remains within one of the two wells. This case also rules out SR manifestation.

\begin{figure}
\[ \includegraphics[scale=0.5]{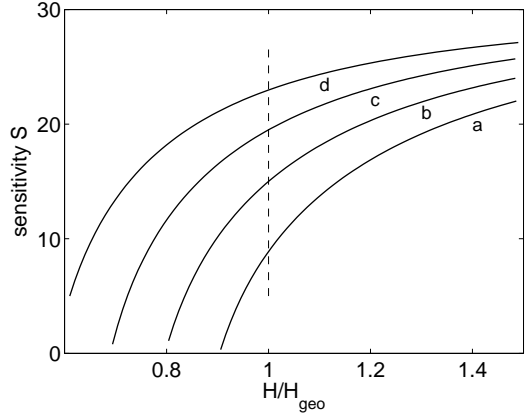} \]
\caption{Sensitivity of the transition probability to MF variations: $k/\mu H_{\rm geo}=$ 0.8 --- a, 0.7 --- b, 0.6 --- c, 0.5 --- d.}
\label{f.3}
\end{figure}

Apparently, for a portion of magnetosomes, large angular chaotic turns take place in the absence of an ac MF also. If some biochemical reaction depends on these turns, it is evident that it must be sensitive to the condition of a `magnetic vacuum' $h\ll H \ll H_{\rm geo}$. Furthermore, the reaction must be sensitive to small variations of the constant MF, since the probability of transition $W$ from well to well {\em exponentially} depends on the barrier height $U_0$, for example in \cite{mcnamara89}:
\[ W=\frac1{2\pi} \sqrt{|U''(0)| U''(\varphi_{\pm})} \exp \left( -2 U_0 / D \right) ~.\] All quantities here, including $U_0$, are functions of the variable $a=k/\mu H$, and hence of $H$. What is of interest is the relative value of the changes in this probability at small variations of the constant MF, i.e., the quantity 
\[ S = -\frac 1W \frac {\textrm{d} W} {\textrm{d} (H/H_{\rm geo})} ~.\] Since the probability drops with the growth of the barrier height, we use `$-$' to hold positive values for the sensitivity $S$. Shown in Fig.\,\ref{f.3} is the sensitivity $S$ computed at several values of the elasticity of the bond between a mean magnetosome and cytoskeleton. It is seen that in a wide range of elasticities the sensitivity of the relative probability to MF variations near $H_{\rm geo}$ is equal to 10--20. This means a 1\% MF change causes 10--20\% changes in the transition probability. Assuming 10\% changes to be biologically significant, we arrive at the limit of detectable values of the constant MF variations $ \sim 0.005 H_{\rm geo}$ or $0.2\mbox{\,$\mu $T}$. This finding generally does not rule out the possibility of a biological system containing magnetosomes to react to slow geomagnetic fluctuations. 

The authors gratefully acknowledge M.M.\,Glaser for improving English style in the article. 

Grants: RFFBR No.04-04-97298, RFH No.04-03-00069.

\end{document}